# Bottom-Up Synthesis of Hexagonal Boron Nitride Nanoparticles with Intensity-Stabilized Quantum Emitters

Yongliang Chen,[1] Xiaoxue Xu,[1] Chi Li,[1] Avi Bendavid,[2] Mika T. Westerhausen,[1] Carlo Bradac,[3] Milos Toth,[1,4] Igor Aharonovich,[1,4] Toan Trong Tran[1,*]

[1]School of Mathematical and Physical Sciences, University of Technology Sydney, Ultimo, NSW, 2007, Australia.

[2]CSIRO Manufacturing, 36 Bradfield Road, Lindfield, New South Wales 2070, Australia.

[3]Trent University, Department of Physics & Astronomy, 1600 West Bank Drive, Peterborough, ON, K9L 0G2

[4]ARC Centre of Excellence for Transformative Meta-Optical Systems (TMOS), University of Technology Sydney, Ultimo, New South Wales 2007, Australia.

*Corresponding author: trongtoan.tran@uts.edu.au

## ABSTRACT

Fluorescent nanoparticles are widely utilized in a large range of nanoscale imaging and sensing applications. While ultra-small nanoparticles (size ≲10 nm) are highly desirable, at this size range their photostability can be compromised due to effects such as intensity fluctuation and spectral diffusion caused by interaction with surface states. In this letter, we demonstrate a facile, bottom-up technique for the fabrication of sub-10-nm hBN nanoparticles hosting photostable bright emitters via a catalyst-free hydrothermal reaction between boric acid and melamine. We also implement a simple stabilization protocol that significantly reduces intensity fluctuation by ~85% and narrows the emission linewidth by ~14% by employing a common sol-gel silica coating process. Our study advances a promising strategy for the scalable, bottom-up synthesis of high-quality quantum emitters in hBN nanoparticles.

## INTRODUCTION

Fluorescent nanoparticles play an integral role in many fields ranging from biomedical imaging and sensing to nanophotonic devices and applications.[1-3] The landscape of fluorescent nanoparticles available for specific realizations is vast and include quantum dots,[4] carbon dots,[5, 6] upconversion nanoparticles,[7] fluorescent silica beads,[8] nano-ruby[9] and nanodiamonds[10-13] to cite a few. For sensing applications based on the optical detection of these nanoparticles, bright and sharp spectral photoluminescence are highly desirable features as they lead to better resolutions. These requirements are usually met by quantum emitters consisting of atom-like defects in nanocrystals.[9, 11] However, the production of these nanoparticles with sizes below 10 nm remains challenging due to their extreme hardness.[14, 15] Quantum emitters in hexagonal boron nitride (hBN)—a two-dimensional, wide-bandgap semiconductor—are thus emerging as a promising alternative. They display ultrahigh brightness, and robustness in harsh environments[16-21] and, unlike diamond and sapphire, the host hBN material allows for top-down fabrication of nanoparticles from powders of micron-sized hBN beads in a straightforward manner.

A few approaches have been developed for breaking down large hBN micro-sized crystals into small hBN nanoparticles. A common technique is ball-milling of substances such as benzyl benzoate where the shear force exerted by the milling balls results in a mixture of hBN nanosheets and nanoparticles.[22] However, the yield of hBN nanoparticles using this technique is significantly lower than that of nanosheets. An alternative approach exploits acoustic cavitation in solvent-mediated ultrasonication processes to break large hBN flakes into nanoparticles.[23, 24] This method typically produces high yields of hBN nanoparticles, yet the nanoparticles obtained via this method are well above 10 nm in size. Recently, a method based on cryogenic-induced cracking of hBN has successfully demonstrated the fabrication of hBN nanoparticles below 10 nm in size, with high-yield.[25, 26] Nevertheless, this top-down approach has two main limitations. i) The size range of the synthesized hBN nanoparticles is relatively large (~5-100 nm).[25] ii) The hosted quantum emitters are highly non-photostable and display photoluminescence intermittency (blinking). The former issue is typical of any top-down approach of this type; in fact, bottom-up routes have been shown to significantly reduce the size distribution of the nanoparticles. The latter problem is more ubiquitous and less method-specific; it is usually related to the increase of surface states associated with, e.g. dangling bonds, point-defects, or trapped charges—whose effect becomes more prominent as the particle size reduces.[14, 27-29] These surface states have been reported to induce fluorescence intensity fluctuation and spectral diffusion because they generate randomly distributed electric fields, which destabilize the optical dipole moments of the emitters via spontaneous Stark shifts.[30-32] One practical solution would be to perform a surface passivation

step—where the surface bonds are chemically terminated to reduce random fluctuations of the local electric field.[13, 14, 32, 33]

In this work, we combine a novel bottom-up synthesis method and surface passivation step to tackle both these issues. The synthesis of the hBN nanoparticles is hydrothermal;[34] it yields particles with diameters below 10 nm with a uniform size distribution. We also demonstrate a facile sol-gel silica coating method that significantly reduces blinking and spectral diffusion of the hosted quantum emitters.

## RESULTS AND DISCUSSION

We adapted a one-pot hydrothermal approach between boric acid and melamine to synthesize hBN nanoparticles.[34] The nanoparticles contain quantum emitters (see below). The synthesis protocol starts with mixing boric acid and melamine in deionized water in a Teflon-lined autoclave reactor for an hour to fully dissolve the two chemicals into an aqueous solution (**Figure 1a**, left panel). The reactor is then heated to 200 °C for 24h in an oven at atmospheric pressure; the reactor is then allowed to cool overnight to room-temperature. More details on the synthesis procedure can be found under the Methods section in the Supporting Information document. The resulting solution is clear and contains small hBN nanoparticles. Note that the nanoparticles can be used directly for characterization without undergoing any additional filtering or centrifugation step. To characterize the size of the hBN particles, we drop-cast the as-synthesized solution on a copper grid with lacey supporting carbon film and performed transmission electron microscopy (TEM) measurements. A representative low-magnification TEM image is shown in **Figure 1b**, with tens of hBN small nanoparticles clearly distinguishable. A size distribution (inset of **Figure 1b**) drawn from a sampling size of 59 particles suggests the average size of the hBN nanoparticles to be ~(6.3 ± 1.1) nm, implying a significantly narrower distribution than that from any top-down methods.[22-26, 35] As shown in **Figure 1c**, the corresponding fast Fourier transform (FFT) image taken from the same area reveals two diffraction rings, associated with the d-spacing of 0.225 nm and 0.206 nm, respectively. These d-spacings translate to the (100) and (101) planes of hexagonal boron nitride lattice structure, respectively.[35, 36] A high-resolution TEM image taken on a representative nanoparticle confirmed the highly crystalline structure of these nanoparticles, with the hBN (100) d-spacing of 0.225 nm (**Figure 1d**). To further confirm the thickness of the nanoparticles, we performed atomic force microscopy (AFM) measurement. A typical AFM image is shown in **Figure 1e**, revealing a thickness of ~9 nm for a typical particle, indicating that the nanoparticles are pseudo-spherical.

Next, we employed X-ray Photoelectron Spectroscopy (XPS) to analyze the chemical composition of the as-synthesized hBN nanoparticles. A survey XPS spectrum (supporting information **Figure S1**) shows the characteristic 1s spectral lines of oxygen (532.5 eV), nitrogen (192.8 eV), carbon (284.4 eV) and boron (402.6 eV). The existence of carbon is primarily due to the carbon contained in melamine used as one of the reactants. A detailed scan around the 1s lines of boron (**Figure 1f**) unveiles a relatively broad peak that could be fitted with three individual peaks at 191.5 eV, 192.3 eV and 193.6 eV, corresponding to B-N (cyan), B-O (green) and B-C (pink) bonds, respectively. The broad peak around the 1s lines of nitrogen (**Figure 1g**) can also be fitted with three individual peaks at 398.2, 400.1 and 402.1 eV, well-matched with the N-B (pink), N-C (cyan) and N-H (green) bonds. These results show good agreement with previous studies, suggesting that hBN nanoparticles fabricated with this method display functional groups on their surface.[34]

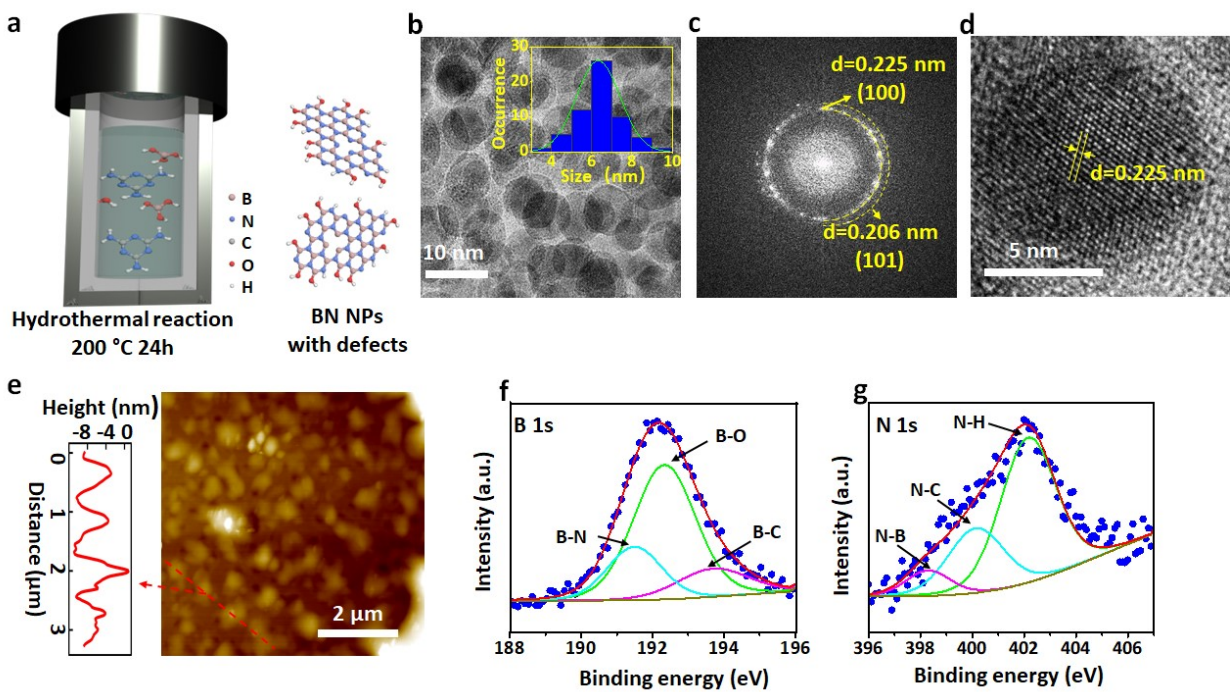

**Figure 1.** Hydrothermal synthesis and characterization of hBN nanoparticles. **(a)** Schematic of the hydrothermal reaction process and the resultant hBN nanoparticles hosting quantum emitters.

Boric acid and melamine are used as precursors. **(b)** TEM image of the hBN nanoparticles. Inset: size distribution statistics indicating an average size of the hydrothermal hBN nanoparticles of ~6.5 nm with a standard deviation of 1.1 nm. **(c)** Representative FFT-TEM image and **(d)** HRTEM image taken from hBN nanoparticles. **(e)** AFM image (right) taken from hBN nanoparticles on Si/SiO$_2$ substrate and height profiles (left) from the red line in the AFM image. **(f)** and **(g)** XPS narrow scan of B 1s and N 1s taken from hBN nanoparticles on Si/SiO$_2$ substrate.

Previous studies established that an annealing process is critically important for the formation of stable, bright emitters in solvent-exfoliated flakes.[18] Full understanding of the role played by the thermal treatment in emitter formation is still lacking. To find the optimal thermal treatment conditions, we conducted two sets of experiments to examine the effect of (i) various gaseous environments and (ii) different temperature. We first examined the effect of annealing with different gaseous environments. Specifically, we kept the annealing temperature constant at 1100 °C for 4 hours and controlled the pressure of the annealing environment to be: 1 Torr of Ar, 1 Torr of O$_2$, or air (760 Torr) with atmospheric pressure. The results are summarized in **Table 1**. Notably, high yield of quantum emitters was only observed when the samples were annealed at 1-2 Torr of O$_2$. Higher O$_2$ pressure resulted in a complete decomposition of the hBN nanoparticles, while lower O$_2$ pressure produced samples with a low yield of single quantum emitters. These results indicate that O$_2$ pressure plays a critical role in producing/activating quantum emitters in hBN nanoparticles. The underlying mechanism of this process is not yet understood and will be further explored in future works.

| Boron precursor | Nitrogen precursor | Hydrothermal synthesis | Annealing condition | Annealing atmosphere | Quantum emitter (QE) photoluminescence |
|---|---|---|---|---|---|
| Boric acid | Melamine | 200 °C, 24 h | 1100 °C for 4h | 1 Torr Ar | •Ensembles<br>•Single QEs<br>(Density:(2.8-5.6) × 10$^2$ /mm$^2$) |

| | | | | 0.5 Torr O$_2$ | • Ensembles<br>• Single QEs<br>(Density: (0.3-1.1) × 10$^3$ /mm$^2$) |
|---|---|---|---|---|---|
| | | | | 1 Torr O$_2$ | • Ensembles<br>• Single QEs<br>(Density: (0.06-2.22) × 10$^5$ /mm$^2$) |
| | | | | 2 Torr O$_2$ | • Ensembles<br>• Single QEs<br>(Density: (0.06-2.22) × 10$^5$ /mm$^2$) |
| | | | | Atmospheric pressure | No emitters |

**Table 1.** Yield of quantum emitters in hBN nanoparticles under different gaseous environments (highlighted in blue) during annealing. Three 60 × 60 μm$^2$ areas were randomly selected for each sample to determine the density of quantum emitters.

Next, we fixed the O$_2$ pressure at 1 Torr, and varied the annealing temperature. The results are shown in **Table 2** and suggest that the annealing temperature needs to be at least 1100 °C to achieve a high yield of quantum emitters. When the annealing temperature is < 1100°C, ensembles of QEs were observed, and we were unable to isolate single quantum emitters. These results led us to choose 1100°C at 1 Torr of O$_2$ as the optimal annealing conditions for the rest of the study. We also observed that samples prepared using either melamine or ammonia in the hydrothermal reaction yielded small (≲10 nm) hBN nanoparticles containing quantum emitters without any detectable difference between the two (**Table 3**). For simplicity, we thus chose melamine as the nitrogen precursor throughout the rest of the experiments.

| Boron precursor | Nitrogen precursor | Hydrothermal reaction time | Annealing condition | Annealing atmosphere | Quantum emitter photoluminescence |
|---|---|---|---|---|---|
| Boric acid | Melamine | 200 °C, 24 h | 1000 °C for 4h | 1 Torr $O_2$ | •Ensembles<br>•Single QEs<br>(Density:(2.8-5.6) × $10^2$ /mm$^2$) |
| | | | 1100 °C for 4h | | •Ensembles<br>•Single QEs<br>(Density:(0.06-2.22) × $10^5$ /mm$^2$) |
| | | | 1200 °C for 4h | | •Ensembles<br>•Single QEs<br>(Density:(0.06-2.22) × $10^5$ /mm$^2$) |
| | | | 1300 °C for 4h | | •Ensembles<br>•Single QEs<br>(Density:(0.06-2.22) × $10^5$ /mm$^2$) |

**Table 2.** Yield of quantum emitters in hBN nanoparticles under different annealing temperatures (highlighted in blue) during annealing. Three 60 × 60 µm² areas were randomly selected for each sample to determine the density of quantum emitters.

| Boron precursor | Nitrogen precursor | Hydrothermal synthesis | Annealing condition | Annealing atmosphere | Quantum emitter photoluminescence |
|---|---|---|---|---|---|
| Boric acid | Melamine | 200 °C, 24h | 1100 °C for 4h | 1 Torr O₂ | •Ensembles<br>•Single QEs<br>(Density:(0.06-2.22) × 10⁵ /mm²) |
| | Ammonia | | | | |

**Table 3.** Yield of quantum emitters in hBN nanoparticles using Melamine or Ammonia as Nitrogen precursors (highlighted in blue) for the hydrothermal synthesis. Three 60 × 60 µm² areas were randomly selected for each sample to determine the density of quantum emitters.

We drop-cast the hBN nanoparticle solution onto a marked 300-nm SiO$_2$/Si substrate and activated the emitters by a thermal annealing step in a tube furnace at 1100 °C for 4 h under 1 Torr of O$_2$. The sample was then characterized using a lab-built micro-photoluminescence (µPL) confocal microscope equipped with a 532-nm CW laser as the excitation source that was focused through a 0.9 numerical aperture (NA) objective lens onto the sample. A dichroic beam splitter and a 550-nm long pass filter were placed in the collection path to reject the back-reflected laser.

**Figure 2a** shows a typical 6 × 6 μm$^2$ confocal map with bright spots that corresponds to the AFM images in **Figure 1e**, indicating high-yield of defects in the hydrothermal-synthesis of hBN nanoparticles. **Figure 2b** shows the histogram of the zero-phonon line (ZPL) wavelength data taken from 115 quantum emitters in the hBN nanoparticles obtained via hydrothermal-synthesis. Notably, the ZPL wavelengths of the quantum emitters mostly distribute at ~580 nm. These result are in agreement with those from previous studies.[37] Unlike the wide ZPL distribution found in typical solvent-exfoliated flakes, in our samples the distribution of ZPLs is narrow, which is advantageous for several applications involving, for instance, integration with external photonic structures.[32, 38, 39] The quantum emitters in our study display very high brightness, with the majority of them exhibiting more than 500 kcounts/s at 2.5 mW of 532-nm excitation with a numerical aperture of 0.9. These count rates are comparable or higher than those of most emitters found in diamond,[14, 28] silicon carbide,[40, 41] rare-earth material,[42] and carbon nanotubes.[43]

Next, we positioned the laser spot onto individual emitters to explore their optical properties. From the confocal map in **Figure 2a**, we found 8 quantum emitters exhibiting single-photon emission (circled in red). Four representative cases, emitter #1, #2, #3 and #4, are shown in **Figure 2c**. The top panels display the spectrum of the various quantum emitters; note the sharp ZPL followed at longer wavelengths by a relatively broad phonon sideband (PSB) associated with the one-phonon replica emission. To verify the quantum emission nature of these emitters, we resorted to second-order autocorrelation measurement, $g^{(2)}(\tau)$, by using a Hanbury Brown and Twiss (HBT) interferometer in collection. All the measurements show strong photon antibunching, with dips at $\tau = 0$ in the rage ~0.1– 0.13 (**Figure 2c**, bottom panel), indicating the single-photon nature of these emitters (conventionally assigned for $g^{(2)}(0) < 0.5$).[44]

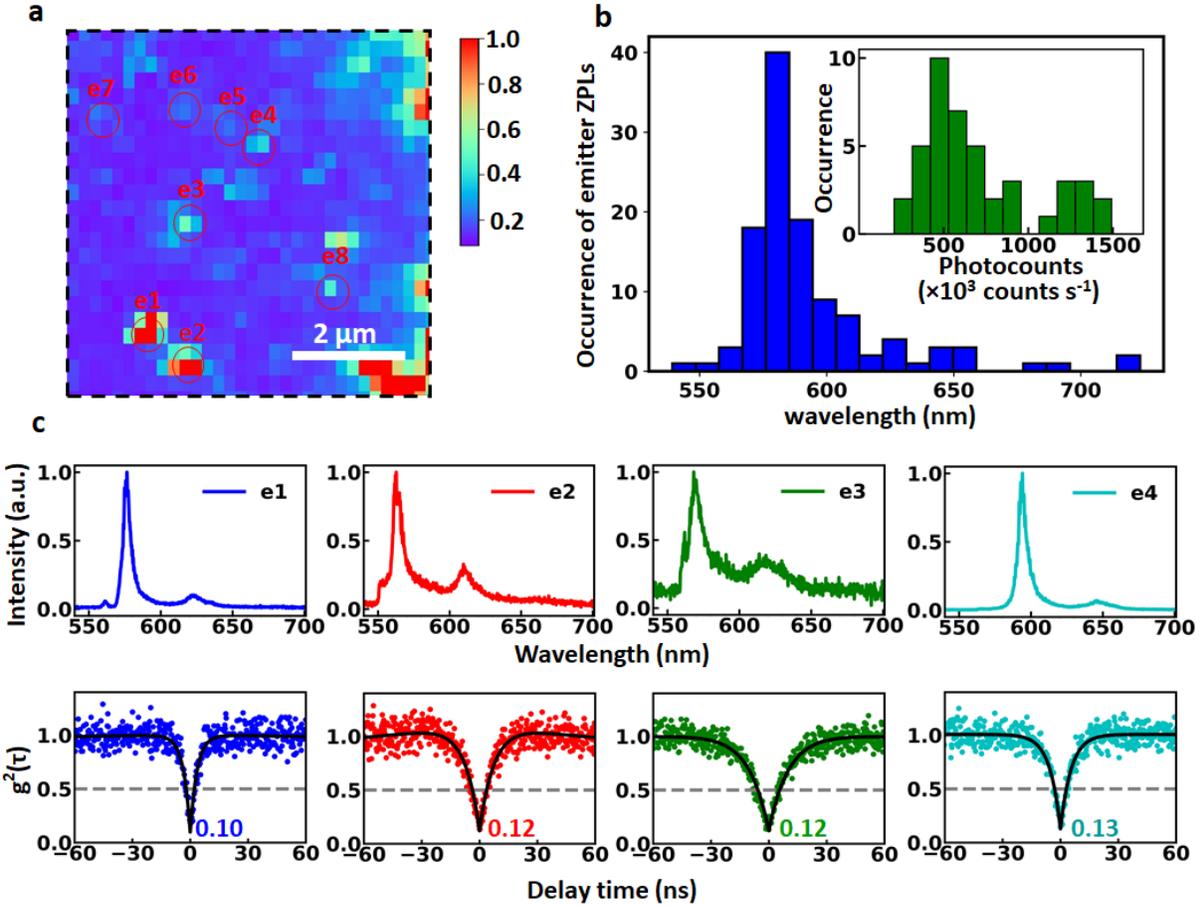

Figure 2. Optical characterization of quantum emitters in hBN nanoparticles synthesized by hydrothermal route. (a) A typical 6 × 6 μm² PL confocal map (the red circles indicate the position of the quantum emitters), corresponding to the area shown in the AFM image in Figure 1d. (b) Histogram of ZPL wavelengths taken from 115 quantum emitters. Inset: histogram of count rates from 45 quantum emitters. The count rates were obtained at 2.5 mW of 532-nm excitation. (c) PL spectrum (top panels) and second order autocorrelation measurement (bottom panels) for four representative quantum emitters (e1–e4) in the scanned region. The PL spectra had an acquisition time of 30 s, at 300 μW and 532-nm excitation, at room temperature. The corresponding autocorrelation measurements were acquired using a 550-nm longpass filter only. The g²(0) of approximately 0.1 indicates the single-photon nature of the quantum emitters.

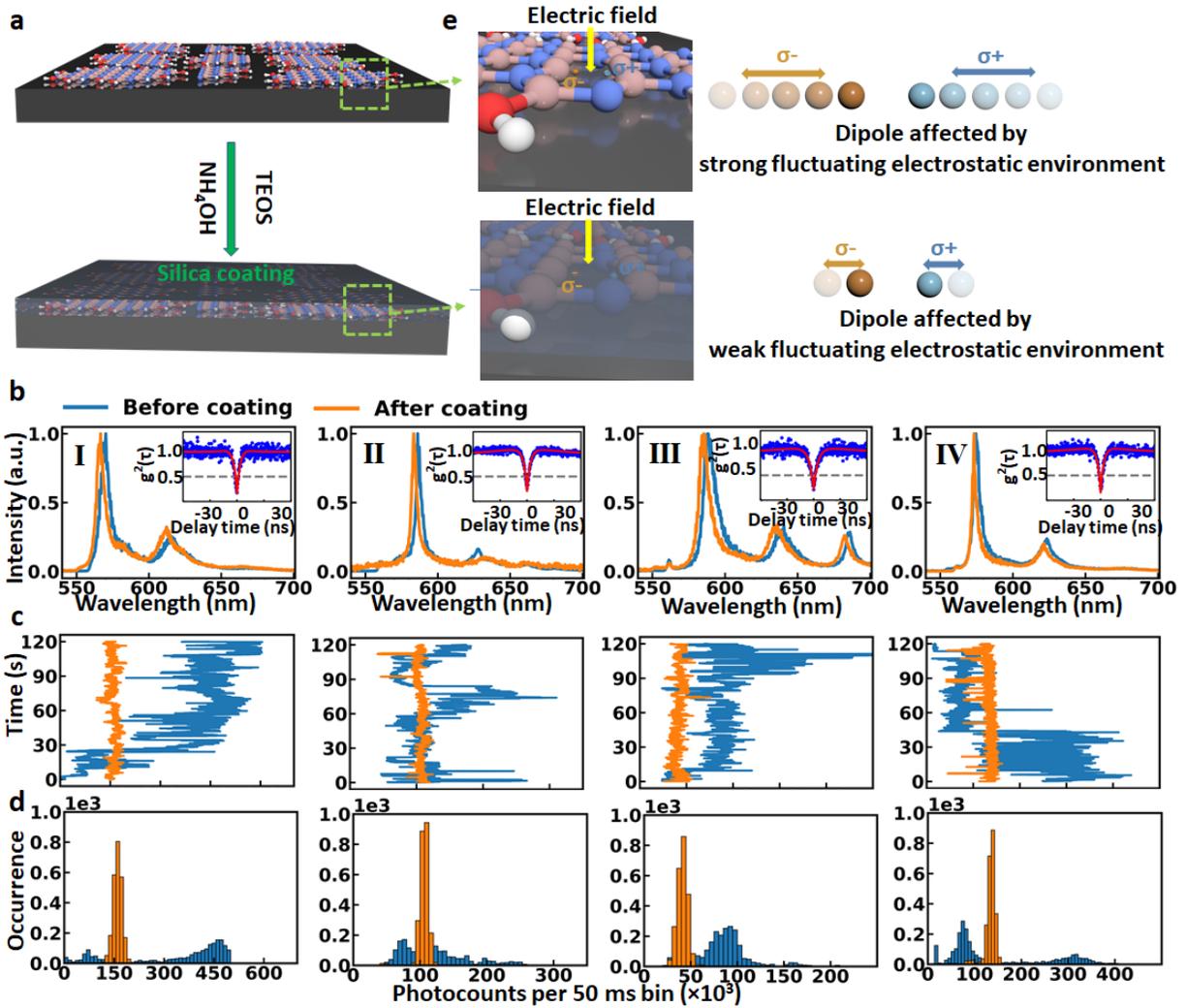

**Figure 3.** Stabilization of blinking emitters via chemical encapsulation of hydrothermal-synthesized hBN nanoparticles. Data plotted for the bare emitters and the silica-coated emitters in blue and orange, respectively. **(a)** Schematic of the silica coating process of the hBN nanoparticles via sol-gel method. **(b)** Stabilization of four representative blinking quantum emitters in hBN nanoparticles, before and after coating (acquisition time of 30 s, at 300 μW, 532-nm excitation, at room temperature). Insets: corresponding second order autocorrelation measurements taken from the four emitters after coating. An acquisition time of 30 s was used. We assume that the emitters only have one single transition state, so we adopted single Lorenzen peak to fit the spectra and found that the full width half maximum (FWHM) is changed from 6.95 nm to 6.30 for emitter I, from 4.97 nm to 4.36 nm for emitter II, from 8.47 nm to 6.97 nm for emitter III and from 4.95 nm to 4.05 nm for emitter IV after silica coating. **(c)** Fluorescence dynamics comparison between the bare and coated emitters. **(d)** Corresponding histograms of photon counts from the four emitters, before and after silica coating. The time bin was 50 ms. **(e)** Schematic describing the hypothesized mechanism for the stabilization induced by silica-coating. Silica coating greatly decreases the fluctuation of the electrostatic environment due to surface states in hBN nanoparticles.

Across the extended survey on several tens of emitters, we observed that most of these exhibit a certain degree of photoluminescence instability (blinking) and spectral diffusion—that can be attributed to the reduced size of the host nanoparticles.[14, 25, 28] As the size of the host particles reduces, the atom-like quantum emitters are statistically more likely to be close to the surface.[27] Such proximity allows surface states such as dangling bonds, trapped charges or functional groups to perturb the dipole moment of the emitters via a phenomenon known as spontaneous Stark shift.[30-32] These inhomogeneous fluctuating electric fields are largely known to be responsible for blinking and spectral diffusion in solid-state quantum emitters—hindering their practical use in applications that require both intensity and spectral stability.[33]

To tackle these detrimental effects, a viable strategy is to passivate the surface states in the hBN nanoparticles. We implemented a simple silica coating method with a common sol-gel route known as the Stober reaction.[45, 46] **Figure 3a** shows a schematic illustrating the hBN nanoparticles before and after silica-coating. For a typical silica coating, a 300-nm $SiO_2$/Si substrate containing annealed hBN nanoparticles were immersed into a mixture of water and ethanol; a portion of tetraethyl orthosilicate (TEOS) was then added so reach a TEOS concentration of ~ 2.9 mM in the final solution.[45, 46] The solution was vigorously stirred to ensure the uniform distribution of the precursor across the reaction solution. Such a silica coating layer is visible through optical microscopy due to the difference in the refractive index between hBN (~1.8) and silica (~1.5), allowing us to observe a thin layer of silica atop the nanoparticle and other areas of the substrate.

To evaluate the effectiveness of the silica coating, we directly compared the optical characteristics of individual emitters in their bare form and after being coated with silica (Figure 3b–d). We randomly chose four representative quantum emitters—whose quantum emissions were verified by the autocorrelation measurements (**Figure 3b**, inset). We performed photoluminescence measurements on the same emitters before and after they were subjected to the silica coating process. **Figure 3b** shows the emission spectra of the bare emitters and the silica-coated ones. The spectra of the bare and silica-coated emitters are almost identical to one another, with slight shifts in wavelengths, likely due to the changes in the dielectric environment surrounding the emitters or to strain, which alter their dipole moments.[13, 14, 32, 33, 43, 47]

To monitor the differences in the photo-dynamics of the emitters, we recorded the emission intensity fluctuation as a function of time—with a 50-ms time bin chosen to enable observation of the blinking behaviors.[48] As shown in **Figure 3c**, while the bare emitters exhibit significant intensity fluctuations, the coated emitters display a substantial improvement in their emission stability—showing an average reduction by ~85% of the emission fluctuations based on the standard deviation calculation. To enable better visualization of the effect, we overlaid the histograms taken from the bare and coated emitters (**Figure 3d**). While the intensity counts of the bare emitters display a bi- or multi-modal distribution of (fluorescence) states, only a single peak is observed for the coated emitters. A bi-/multi-modal distribution of the intensity counts typically implies the existence of two or more states of the system.[14, 29, 49, 50] The existence of multitude emission is problematic for practical applications.[33]

Next, we investigated whether spectral diffusion was responsible for the difference in linewidths between bare and coated emitters. To allow for such a comparison, we fitted the ZPLs of the bare and coated emitters with a single Lorentzian peak (i.e. assuming a single transition dipole from the emitters). The silica-coating resulted in a narrowing of the linewidths from 6.95 nm to 6.30 for emitter I, from 4.97 nm to 4.36 nm for emitter II, from 8.47 nm to 6.97 nm for emitter III, and from 4.95 nm to 4.05 nm for emitter IV. On average, the linewidths were reduced by ~14% after the silica coating process. Since the measurement temperature and excitation power were kept constant for both the measurements on bare and coated hBN nanoparticles, the linewidth contribution from the phonon-induced broadening of the ZPLs was the same for the two cases. As such, the decrease in the linewidths of ZPLs from the coated nanoparticles suggests that there was a considerable reduction in spectral diffusion from these emitters.[43, 50] From these results, we hypothesize that the silica-coating film acts as a passivating layer that neutralizes the surface states such as dangling bonds, functional groups, surface point defects, etc. Such a passivating effect, in turn, reduces the spontaneous Stark effects caused by the optical cycling of these surface states (**Figure 3e**).[30, 32]

In conclusion, we have demonstrated a hydrothermal bottom-up synthesis of quantum emitters in hBN nanoparticles. The quantum emitters were of high brightness and show good single-photon purity. We have also established a facile sol-gel method to coat a thin silica film onto the hBN nanoparticles and have shown that such a protocol significantly improves the optical properties of the emitters—narrowing their linewidths and reducing intensity fluctuations. Our study lays the foundation towards the fabrication of spectrally-stabilized quantum emitters in hBN ultra-small nanoparticles (≲10 nm) for a range of applications such as bioimaging and nanothermometry, among many others.

## Associated Content

### Corresponding author

Toan Trong Tran - School of Mathematical and Physical Sciences, University of Technology Sydney, Ultimo, NSW, 2007, Australia; orcid.org/0000-0001-8960-0253; Email: trongtoan.tran@uts.edu.au

### Authors

Yongliang Chen - School of Mathematical and Physical Sciences, University of Technology Sydney, Ultimo, NSW, 2007, Australia.

Xiaoxue Xu - School of Mathematical and Physical Sciences, University of Technology Sydney, Ultimo, NSW, 2007, Australia.

Chi Li - School of Mathematical and Physical Sciences, University of Technology Sydney, Ultimo, NSW, 2007, Australia

Avi Bendavid - CSIRO Manufacturing, 36 Bradfield Road, Lindfield, New South Wales 2070, Australia.

Mika T. Westerhausen - School of Mathematical and Physical Sciences, University of Technology Sydney, Ultimo, NSW, 2007, Australia.

Carlo Bradac - [3]Trent University, Department of Physics & Astronomy, 1600 West Bank Drive, Peterborough, ON, K9L 0G2; orcid.org/0000-0002-6673-7238

Milos Toth - School of Mathematical and Physical Sciences, University of Technology Sydney, Ultimo, NSW, 2007, Australia; and

ARC Center of Excellence for Transformative Meta-Optical Systems (TMOS), Faculty of Science, University of Technology Sydney, Australia; orcid.org/0000-0003-1564-4899.

Igor Aharonovich - School of Mathematical and Physical Sciences, University of Technology Sydney, Ultimo, NSW, 2007, Australia; and

ARC Center of Excellence for Transformative Meta-Optical Systems (TMOS), Faculty of Science, University of Technology Sydney, Australia; orcid.org/0000-0003-4304-3935.

## Author Contributions

The project was conceived and supervised by T. T. T. The experiments were carried out by Y. C. with the assistance of C. L., C. B., X. X., and M. W. The XPS characterization and analysis was carried out by A. B. The TEM characterization and analysis was conducted by X.X. All the authors contributed to analyzing the data and writing the paper.

## Notes

The authors declare no competing financial interest.

## Fundings

We acknowledge financial support from the Australian Research Council (via DP180100077, and DP190101058).

## Acknowledgments

We thank Dr. Alexander Solntsev, Dr. Johannes Froch and Dr. Mehran Kianinia for the fruitful discussions and Mr. Thinh N. Tran for his technical assistance in data analysis.

## References

(1) Wolfbeis, O. S., An Overview of Nanoparticles Commonly Used in Fluorescent Bioimaging. *Chem Soc Rev* **2015,** *44*, 4743-68.

(2) Zhang, J.; Cheng, F.; Li, J.; Zhu, J. J.; Lu, Y., Fluorescent Nanoprobes for Sensing and Imaging of Metal Ions: Recent Advances and Future Perspectives. *Nano Today* **2016,** *11*, 309-329.

(3) Panfil, Y. E.; Oded, M.; Banin, U., Colloidal Quantum Nanostructures: Emerging Materials for Display Applications. *Angew Chem Int Ed Engl* **2018,** *57*, 4274-4295.

(4) Hanifi, D. A.; Bronstein, N. D.; Koscher, B. A.; Nett, Z.; Swabeck, J. K.; Takano, K.; Schwartzberg, A. M.; Maserati, L.; Vandewal, K.; van de Burgt, Y.; Salleo, A.; Alivisatos, A. P., Redefining near-Unity Luminescence in Quantum Dots with Photothermal Threshold Quantum Yield. *Science* **2019,** *363*, 1199-1202.

(5) Liu, J. H.; Cao, L.; LeCroy, G. E.; Wang, P.; Meziani, M. J.; Dong, Y.; Liu, Y.; Luo, P. G.; Sun, Y. P., Carbon "Quantum" Dots for Fluorescence Labeling of Cells. *ACS Appl Mater Interfaces* **2015,** *7*, 19439-45.

(6) Wang, X.; Feng, Y.; Dong, P.; Huang, J., A Mini Review on Carbon Quantum Dots: Preparation, Properties, and Electrocatalytic Application. *Front Chem* **2019,** *7*, 671.

(7) Qiu, X.; Zhou, Q.; Zhu, X.; Wu, Z.; Feng, W.; Li, F., Ratiometric Upconversion Nanothermometry with Dual Emission at the Same Wavelength Decoded Via a Time-Resolved Technique. *Nature Communications* **2020,** *11*.

(8) Sun, W.; Sun, Q.; Zhao, Q.; Marin, L.; Cheng, X., Fluorescent Porous Silica Microspheres for Highly and Selectively Detecting Hg(2+) and Pb(2+) Ions and Imaging in Living Cells. *ACS Omega* **2019,** *4*, 18381-18391.

(9) Edmonds, A. M.; Sobhan, M. A.; Sreenivasan, V. K. A.; Grebenik, E. A.; Rabeau, J. R.; Goldys, E. M.; Zvyagin, A. V., Nano-Ruby: A Promising Fluorescent Probe for Background-Free Cellular Imaging. *Particle & Particle Systems Characterization* **2013,** *30*, 506-513.

(10) Tran, T. T.; Regan, B.; Ekimov, E. A.; Mu, Z.; Zhou, Y.; Gao, W.-b.; Narang, P.; Solntsev, A. S.; Toth, M.; Aharonovich, I.; Bradac, C., Anti-Stokes Excitation of Solid-State Quantum Emitters for Nanoscale Thermometry. *Science Advances* **2019,** *5*, eaav9180.

(11) Sotoma, S.; Epperla, C. P.; Chang, H.-C., Diamond Nanothermometry. *ChemNanoMat* **2018,** *4*, 15-27.

(12) Kucsko, G.; Maurer, P. C.; Yao, N. Y.; Kubo, M.; Noh, H. J.; Lo, P. K.; Park, H.; Lukin, M. D., Nanometre-Scale Thermometry in a Living Cell. *Nature* **2013**, *500*, 54.

(13) Nagl, A.; Hemelaar, S. R.; Schirhagl, R., Improving Surface and Defect Center Chemistry of Fluorescent Nanodiamonds for Imaging Purposes--a Review. *Anal Bioanal Chem* **2015**, *407*, 7521-36.

(14) Bradac, C.; Gaebel, T.; Naidoo, N.; Sellars, M. J.; Twamley, J.; Brown, L. J.; Barnard, A. S.; Plakhotnik, T.; Zvyagin, A. V.; Rabeau, J. R., Observation and Control of Blinking Nitrogen-Vacancy Centres in Discrete Nanodiamonds. *Nat Nanotechnol* **2010**, *5*, 345-9.

(15) Fabrication and Characterization of Ruby Nanoparticles. *Malaysian Journal of Analytical Science* **2018**, *22*.

(16) Tran, T. T.; Bray, K.; Ford, M. J.; Toth, M.; Aharonovich, I., Quantum Emission from Hexagonal Boron Nitride Monolayers. *Nature Nanotechnology* **2015**, *11*, 37.

(17) Kianinia, M.; Regan, B.; Tawfik, S. A.; Tran, T. T.; Ford, M. J.; Aharonovich, I.; Toth, M., Robust Solid-State Quantum System Operating at 800 K. *ACS Photonics* **2017**, *4*, 768-773.

(18) Tran, T. T.; Elbadawi, C.; Totonjian, D.; Lobo, C. J.; Grosso, G.; Moon, H.; Englund, D. R.; Ford, M. J.; Aharonovich, I.; Toth, M., Robust Multicolor Single Photon Emission from Point Defects in Hexagonal Boron Nitride. *ACS Nano* **2016**, *10*, 7331-7338.

(19) Hayee, F.; Yu, L.; Zhang, J. L.; Ciccarino, C. J.; Nguyen, M.; Marshall, A. F.; Aharonovich, I.; Vuckovic, J.; Narang, P.; Heinz, T. F.; Dionne, J. A., Revealing Multiple Classes of Stable Quantum Emitters in Hexagonal Boron Nitride with Correlated Optical and Electron Microscopy. *Nat Mater* **2020**, *19*, 534-539.

(20) Gottscholl, A.; Kianinia, M.; Soltamov, V.; Orlinskii, S.; Mamin, G.; Bradac, C.; Kasper, C.; Krambrock, K.; Sperlich, A.; Toth, M.; Aharonovich, I.; Dyakonov, V., Initialization and Read-out of Intrinsic Spin Defects in a Van Der Waals Crystal at Room Temperature. *Nat Mater* **2020**, *19*, 540-545.

(21) Chen, Y.; Tran, T. N.; Duong, N. M. H.; Li, C.; Toth, M.; Bradac, C.; Aharonovich, I.; Solntsev, A.; Tran, T. T., Optical Thermometry with Quantum Emitters in Hexagonal Boron Nitride. *ACS Appl Mater Interfaces* **2020**, *12*, 25464-25470.

(22) Deepika; Li, L. H.; Glushenkov, A. M.; Hait, S. K.; Hodgson, P.; Chen, Y., High-Efficient Production of Boron Nitride Nanosheets Via an Optimized Ball Milling Process for Lubrication in Oil. *Sci Rep* **2014**, *4*, 7288.

(23) Bari, R.; Parviz, D.; Khabaz, F.; Klaassen, C. D.; Metzler, S. D.; Hansen, M. J.; Khare, R.; Green, M. J., Liquid Phase Exfoliation and Crumpling of Inorganic Nanosheets. *Phys Chem Chem Phys* **2015,** *17*, 9383-93.

(24) Smith, R. J.; King, P. J.; Lotya, M.; Wirtz, C.; Khan, U.; De, S.; O'Neill, A.; Duesberg, G. S.; Grunlan, J. C.; Moriarty, G.; Chen, J.; Wang, J.; Minett, A. I.; Nicolosi, V.; Coleman, J. N., Large-Scale Exfoliation of Inorganic Layered Compounds in Aqueous Surfactant Solutions. *Adv Mater* **2011,** *23*, 3944-8.

(25) Duong, N. M. H.; Glushkov, E.; Chernev, A.; Navikas, V.; Comtet, J.; Nguyen, M. A. P.; Toth, M.; Radenovic, A.; Tran, T. T.; Aharonovich, I., Facile Production of Hexagonal Boron Nitride Nanoparticles by Cryogenic Exfoliation. *Nano Letters* **2019**.

(26) Wang, Y.; Liu, Y.; Zhang, J.; Wu, J.; Xu, H.; Wen, X.; Zhang, X.; Tiwary, C. S.; Yang, W.; Vajtai, R.; Zhang, Y.; Chopra, N.; Odeh, I. N.; Wu, Y.; Ajayan, P. M., Cryo-Mediated Exfoliation and Fracturing of Layered Materials into 2d Quantum Dots. *Science Advances* **2017,** *3*, e1701500.

(27) Bradac, C.; Gaebel, T.; Naidoo, N.; Rabeau, J. R.; Barnard, A. S., Prediction and Measurement of the Size-Dependent Stability of Fluorescence in Diamond over the Entire Nanoscale. *Nano Letters* **2009,** *9*, 3555-3564.

(28) Vlasov, II; Shiryaev, A. A.; Rendler, T.; Steinert, S.; Lee, S. Y.; Antonov, D.; Voros, M.; Jelezko, F.; Fisenko, A. V.; Semjonova, L. F.; Biskupek, J.; Kaiser, U.; Lebedev, O. I.; Sildos, I.; Hemmer, P. R.; Konov, V. I.; Gali, A.; Wrachtrup, J., Molecular-Sized Fluorescent Nanodiamonds. *Nat Nanotechnol* **2014,** *9*, 54-8.

(29) Elke Neu, M. A., and Christoph Becher,, Photophysics of Single Silicon Vacancy Centers in Diamond: Implications for Single Photon Emission.Pdf>. *Opt. Express* **2012,** *201*, 19956-1997.

(30) Wolters, J.; Sadzak, N.; Schell, A. W.; Schroder, T.; Benson, O., Measurement of the Ultrafast Spectral Diffusion of the Optical Transition of Nitrogen Vacancy Centers in Nano-Size Diamond Using Correlation Interferometry. *Phys Rev Lett* **2013,** *110*, 027401.

(31) Neuhauser, R. G.; Shimizu, K. T.; Woo, W. K.; Empedocles, S. A.; Bawendi, M. G., Correlation between Fluorescence Intermittency and Spectral Diffusion in Single Semiconductor Quantum Dots. *Physical Review Letters* **2000,** *85*, 3301-3304.

(32) Noh, G.; Choi, D.; Kim, J. H.; Im, D. G.; Kim, Y. H.; Seo, H.; Lee, J., Stark Tuning of Single-Photon Emitters in Hexagonal Boron Nitride. *Nano Lett* **2018,** *18*, 4710-4715.

(33) Wang, X.; Ren, X.; Kahen, K.; Hahn, M. A.; Rajeswaran, M.; Maccagnano-Zacher, S.; Silcox, J.; Cragg, G. E.; Efros, A. L.; Krauss, T. D., Non-Blinking Semiconductor Nanocrystals. *Nature* **2009,** *459*, 686-689.

(34) Huo, B.; Liu, B.; Chen, T.; Cui, L.; Xu, G.; Liu, M.; Liu, J., One-Step Synthesis of Fluorescent Boron Nitride Quantum Dots Via a Hydrothermal Strategy Using Melamine as Nitrogen Source for the Detection of Ferric Ions. *Langmuir* **2017,** *33*, 10673-10678.

(35) Lin, L.; Xu, Y.; Zhang, S.; Ross, I. M.; Ong, A. C.; Allwood, D. A., Fabrication and Luminescence of Monolayered Boron Nitride Quantum Dots. *Small* **2014,** *10*, 60-5.

(36) Hou, L.; Gao, F.; Sun, G.; Gou, H.; Tian, M., Synthesis of High-Purity Boron Nitride Nanocrystal at Low Temperatures. *Crystal Growth & Design* **2007,** *7*, 535-540.

(37) Mendelson, N.; Xu, Z.-Q.; Tran, T. T.; Kianinia, M.; Scott, J.; Bradac, C.; Aharonovich, I.; Toth, M., Engineering and Tuning of Quantum Emitters in Few-Layer Hexagonal Boron Nitride. *ACS Nano* **2019,** *13*, 3132-3140.

(38) Xue, Y.; Wang, H.; Tan, Q.; Zhang, J.; Yu, T.; Ding, K.; Jiang, D.; Dou, X.; Shi, J. J.; Sun, B. Q., Anomalous Pressure Characteristics of Defects in Hexagonal Boron Nitride Flakes. *ACS Nano* **2018,** *12*, 7127-7133.

(39) Chakraborty, C.; Goodfellow, K. M.; Dhara, S.; Yoshimura, A.; Meunier, V.; Vamivakas, A. N., Quantum-Confined Stark Effect of Individual Defects in a Van Der Waals Heterostructure. *Nano Lett* **2017,** *17*, 2253-2258.

(40) Wolfowicz, G.; Anderson, C. P.; Diler, B.; Poluektov, O. G.; Heremans, F. J.; Awschalom, D. D., Vanadium Spin Qubits as Telecom Quantum Emitters in Silicon Carbide. *Science Advances* **2020,** *6*, eaaz1192.

(41) Wang, J.; Zhou, Y.; Wang, Z.; Rasmita, A.; Yang, J.; Li, X.; von Bardeleben, H. J.; Gao, W., Bright Room Temperature Single Photon Source at Telecom Range in Cubic Silicon Carbide. *Nat Commun* **2018,** *9*, 4106.

(42) Kolesov, R.; Xia, K.; Reuter, R.; Stohr, R.; Zappe, A.; Meijer, J.; Hemmer, P. R.; Wrachtrup, J., Optical Detection of a Single Rare-Earth Ion in a Crystal. *Nat Commun* **2012,** *3*, 1029.

(43) Ai, N.; Walden-Newman, W.; Song, Q.; Kalliakos, S.; Strauf, S., Suppression of Blinking and Enhanced Exciton Emission from Individual Carbon Nanotubes. *ACS Nano* **2011,** *5*, 2664-2670.

(44) Li, C.; Xu, Z.-Q.; Mendelson, N.; Kianinia, M.; Toth, M.; Aharonovich, I., Purification of Single-Photon Emission from Hbn Using Post-Processing Treatments. *Nanophotonics* **2019,** *0*.

(45) Kobayashi, Y.; Inose, H.; Nakagawa, T.; Gonda, K.; Takeda, M.; Ohuchi, N.; Kasuya, A., Control of Shell Thickness in Silica-Coating of Au Nanoparticles and Their X-Ray Imaging Properties. *J Colloid Interface Sci* **2011,** *358*, 329-33.

(46) Deng, W.; Jin, D.; Drozdowicz-Tomsia, K.; Yuan, J.; Wu, J.; Goldys, E. M., Ultrabright Eu–Doped Plasmonic Ag@Sio2 Nanostructures: Time-Gated Bioprobes with Single Particle Sensitivity and Negligible Background. *Advanced Materials* **2011,** *23*, 4649-4654.

(47) Mendelson, N.; Doherty, M.; Toth, M.; Aharonovich, I.; Tran, T. T., Strain-Induced Modification of the Optical Characteristics of Quantum Emitters in Hexagonal Boron Nitride. *Adv Mater* **2020,** *32*, e1908316.

(48) Kuno, M.; Fromm, D. P.; Hamann, H. F.; Gallagher, A.; Nesbitt, D. J., "On"/"Off" Fluorescence Intermittency of Single Semiconductor Quantum Dots. *The Journal of Chemical Physics* **2001,** *115*, 1028-1040.

(49) Walden-Newman, W.; Sarpkaya, I.; Strauf, S., Quantum Light Signatures and Nanosecond Spectral Diffusion from Cavity-Embedded Carbon Nanotubes. *Nano Lett* **2012,** *12*, 1934-41.

(50) Tran, T. T.; Bradac, C.; Solntsev, A. S.; Toth, M.; Aharonovich, I., Suppression of Spectral Diffusion by Anti-Stokes Excitation of Quantum Emitters in Hexagonal Boron Nitride. *Applied Physics Letters* **2019,** *115*.